\documentstyle[aps,multicol,epsf]{revtex}
\begin{document}
\draft

\title{Avalanches in the lung: A statistical mechanical model}

\author{ 
Albert-L\'aszl\'o Barab\'asi$^{1,2}$, Sergey V. Buldyrev$^1$, H.
Eugene Stanley,$^1$ and B\'ela Suki$^3$}
\address{
 $^1$Center for Polymer Studies and Department of
Physics, Boston University, Boston, MA 02215
\\$^2$Department of 
Physics,
University
of Notre Dame, Notre Dame, IN 46556
\\ $^3$Respiratory Research Laboratory, Department of
Biomedical Engineering, Boston University, Boston, MA 02215}
\date{LC5777, submitted 6 March 1995}
\maketitle

\begin{abstract}

We study a statistical mechanical model for the dynamics of lung
inflation which incorporates recent experimental observations on
the
opening of individual airways by a cascade or avalanche mechanism.
Using an exact mapping of the avalanche problem onto percolation on
a
Cayley tree, we analytically derive the exponents describing the
size
distribution of the first avalanches and test the analytical
solution by
numerical simulations.  We find that the tree-like structure of the
airways together with the simplest assumptions concerning opening
threshold pressures of each airway, is sufficient to explain the
existence of power-law distributions observed experimentally.

\end{abstract}

\bigskip

\begin{multicols}{2}
 Recent interactions between physics and physiology have resulted
in
advances in understanding some ``simpler'' physiological
systems\cite{grosberg94}.  In particular, considerable progress has
occurred in the general area of statistical mechanics and pulmonary
physiology\cite{Shl91,west89} due most likely to the unique
tree-like
connectivity of the airways\cite{Horsf}. 

During a forced exhalation, lungs deflate to very low volumes and
many
peripheral airways close up \cite{Eng75}.  In lung disease, closure
occurs even during normal breathing; the closed airways do not
reopen
for a significant portion of the following inhalation\cite{Cra89}. 
As a
consequence, a large portion of the alveolar space can remain
closed
during the entire breathing cycle leading to severe hypoventilation
and
an imbalance between ventilation and perfusion.  The process of
opening
a {\it single\/} airway is a local and isolated phenomenon. 
However,
the dynamics of {\it consecutive\/} airway openings in the lung is
a
highly cooperative process.  There is recent evidence suggesting
that
during inflation the resistance to airflow of the small airways
decreases in discrete jumps\cite{otis,Suki94}.  Thus airways
do not open individually, but in a sequence of bursts or
``avalanches''\cite{aval} involving many airways; both the sizes of
these jumps and the time intervals between jumps follow power law
distributions\cite{Suki94}.  In this paper, we argue that the
existence
of power laws in lung inflation can arise directly from the
tree-like
connectivity of the airways.  We observe that the dynamics of lung
inflation can be usefully described by a percolation problem on a
Cayley
tree, with the inflated lung volume corresponding to a percolation
cluster.  Using this exact mapping, we analytically derive the
exponents
describing the size distribution of the {\it first} avalanches, and
test
our results using simulations.

 Morphological data \cite{Horsf} show that human (as well as other
mammalian) lungs constitute an asymmetric branching airway
structure
with $\approx 35$ generations.  Complete airway closure appears to
occur
only in the last $\approx 10-14$ generations \cite{Suki94}, where
the
branching structure is largely symmetric \cite{Horsf}. 
Accordingly, we
model this part of the airway tree as a binary Cayley tree with
airway
segments that can be either closed or opened (Fig. 1).  At time
$t=0$,
all airways are assumed to be closed.  Lung inflation is simulated
by
applying an external pressure $P_E$ at the root of the tree, and
gradually increasing $P_E$ at a uniform and slow rate.  Airways are
labeled $(i,j)$ with a generation number $i$ ($i=0,\ldots,N$),
where $N$
is the order of the tree ($i=0$ denotes the tree root), and a
column
number $j$ ($j=0,\ldots,2^i-1$).  An {\it opening threshold
pressure}
$P_{ij}$ is also assigned to each airway ($i,j$).  Experiments on
flexible tube airway models\cite{Gav90} confirm that the opening of
a
single airway is a dynamic process, with each airway characterized
by a
critical pressure threshold such that if $P_E$ exceeds this
threshold,
then the airway opens in a short time, which is considered to be
instantaneous\cite{note2}.  Opening of airway ($i,j$) occurs
whenever
$P_{ij}$ is smaller than the pressure in its parent.

  We assume that $P_{ij}$ is uniformly distributed between 0 and 1
\cite{distr}, and allow $P_{\rm E}$ to increase from 0 to 1 in
small
increments.  When $P_{\rm E}$ first exceeds $P_{00}$, the airway
($0,0$)
opens and its pressure is set equal to $P_{\rm E}$. Next, the two
airways (1,0) and (1,1) are tested to see if they can be opened
with this value
of $P_{\rm E}$---i.e., if $P_{\rm E}>P_{10}$ and/or $P_{\rm
E}>P_{11}$.
If one or both conditions are met, then the airways ($1,0$) and/or
($1,1$) are also opened. This opening is then continued
sequentially
down the tree until no airway is found with its $P_{ij}<P_{\rm E}$.

Of particular interest is the fact that a small increase in $P_{\rm
E}$
can lead to an ``avalanche" in which many airways open
simultaneously.
When the first avalanche stops, $P_E$ is further incremented and
pressures in the open airways are updated.  We iterate this process
until all airways open.   The location and size of the next
avalanche depends on the distribution of $P_{ij}$ in the accessible
region.

We do not treat the full problem analytically, but we can obtain
exact
results for the distribution of the {\it first\/} avalanche.  At
$t=0$,
we increase $P_E$ until the first avalanche occurs and we calculate
its
size $s$.  Then we restart the simulation with a new set of
thresholds
$\{P_{ij}\}$.

Before we consider two possible definitions of $s$, we note that
gas
exchange in the lung occurs only in the ``opened" alveoli (the
terminal
units of the bronchial tree) which are in communication with the
trachea.  For this reason, in Definition A, $s$ denotes the number
of
alveoli, defined as the number of elements in the last generation,
$N$,
that become open.  Motivated by percolation
theory\cite{Essam80,bunde1},
in Definition B, $s$ is the number airways that open following an
increase of $P_E$ that opens at least one airway.  The
physiological
rationale for definition B is that when the lung is deflated to low
volumes, most airways close.  However, often there remains trapped
air
in the alveoli.  Thus concerning gas exchange it may not be
necessary
that an avalanche reach the bottom of the tree for it to connect
alveoli
with the trachea.

We study $\Pi(s)$, the size distribution of first avalanches.  For
Definition A, $\Pi^A(s)$ shows a single power law behavior with an
exponent $\gamma^A= 0.9$ $(\approx 1.0)$ (Fig.  2).  For Definition
B
the function $\Pi^B(s)$ has two regions (Fig.~2): a first region
with
a steep power-law decay and a second region with a moderate
power-law
decay, with a crossover at a scale $N$,
\begin{equation}
\label{e.1}
\Pi^B(s)\sim \cases{ s^{-\gamma^B_1} &~~~~ $[s \leq N]$ \cr
                         s^{-\gamma^B_2} &~~~~ $[s \gg N]$}.
\end{equation}
The exponent ${\gamma^B_1} = 1.9$ $(\approx 2.0)$ for the first
regime,
while ${\gamma^B_2} = 0.9$ $(\approx 1.0)$ for the second regime
which
extends to sizes including all branches, i.e., almost to a size of
$2^{N+1}-1$.

 We argue that for Definition B, this problem can be mapped onto
the
percolation problem for the Cayley tree\cite{bunde1}.  In the
percolation problem, we occupy randomly every branch of the tree
with a
probability $p$.  Then, starting from the root, we connect all
occupied
branches that are neighbors of each other.  Definition B concerns
the
cluster of connected bonds that starts from the root.  The size of
this
cluster depends on the fraction $p$ of occupied bonds.  As we
approach a
critical probability $p_c$, the typical size of a cluster can be
characterized by the $s_{\rm typ}\sim |p-p_c|^{-1/\sigma}$.  Both
$\sigma$ and $p_c$ can be calculated exactly due to the branching
nature
of the tree: $\sigma={1\over 2}$ and $p_c={1\over 2}$
\cite{bunde1}.  In general, the size distribution of the finite
clusters in the infinite system obeys the scaling form\cite{bunde1}
\begin{equation}
\label{e.2}
\Pi(s) = s^{-\tau} f(s^\sigma |p-p_c|),
\end{equation}
where $\tau=3/2$, and $f(u)=$ const for $u \ll 1$ and $f(u \gg 1)
\to 0$.
 To connect percolation theory to the lung model, instead of
occupying
randomly the branches with probability $p$, we assign a random
number or
pressure threshold value to each airway.  We then define a cluster
to be
the set of airways that have a threshold smaller than a predefined
value
$p$ and are connected to the root.  When $P_E$ exceeds $P_{00}$, we
open
all airways below the root which have a threshold value smaller
than
$P_{00}$.

  If $P_{00}$ is fixed and set equal to $p$, then this is {\it
exactly
the percolation problem on the Cayley tree} and the distribution of
the
cluster sizes or avalanches is given by (\ref{e.2}).  However, in
our
case $P_{00}$ is also a random variable.  Thus, in order to obtain
the
size distribution of the first avalanche, we must integrate the
cluster
distribution over the probability $p$, from 0 to 1 with the result
that 
\begin{equation}
\gamma_1^B=\tau+\sigma,
\end{equation}
which predicts $\gamma_1^B=2$, in agreement with the scaling
observed
for $s \le N$ in Fig. 2.

These calculations assume that the system size is infinite.  No
avalanche with $s<N$ can reach the bottom of the tree, so the
scaling
behavior for $s<N$ is that of the infinite tree with an exponent
$\gamma_1^B=2$.  On the other hand, avalanches of size $s > N$ are
affected by finite size effects, and indeed the data for $s \gg N$
indicate a different exponent.  Moreover, finite-size effects will
always affect the scaling behavior of $\Pi^A(s)$, since every
avalanche
that leads to the opening of one or more alveoli must open at least
$N$
airways.

Percolation theory on the Cayley tree does not provide the scaling
behavior of the cluster size distribution dominated by finite size
effects. Nevertheless, we can obtain the scaling exponent using a
generating function approach, developed in the theory of branching
processes \cite{Harris}.  We define the generating function of
order $N$
\begin{equation}
g_N^{A,B}(p,x)=\sum_{s=0}^\infty P_N^{A,B}(p,s)x^s,
\end{equation}
where $P_N^{A,B}(p,s)$ give, for definitions A and B, the
probability
that in a tree with $N$ generations we have an avalanche of size
$s$ for
a given $P_{00}=p$.  Therefore,
\begin{equation}
\label{e.6}
\Pi_N^{A,B}(s) = \int_0^1 P_N^{A,B}(p,s) dp.
\end{equation}
These generating functions satisfy the recursion relations
\begin{equation}
g_{N+ 1}^{A,B}(p,x) = x^{\phi} [(1-p)+pg_N^{A,B}(p,x)]^2,~~~
g_0^{A,B}(p,x)=x
\end{equation}
where $\phi=0$ for definition A and $\phi=1$ for Definition B,
$g_N^A$
is a polynomial in $x$ of degree $2^N$ and $g_N^B$ is a polynomial
of
degree $2^{N+1}-1$. We obtain the distribution functions $\Pi^A(s)$
and
$\Pi^B(s)$ by numerical integration of the coefficients of these
polynomials with respect to $p=P_{00}$.  The results for $\Pi^A(s)$
and
$\Pi^B(s)$ are shown in Fig.~2; note the good agreement between
simulations and theory, despite the fact that there are no
adjustable
parameters in the calculation (the theoretical line being
determined
solely by the value of $N$).

For $N\to\infty$, for any $x<1$, the generating function
$g^B_N(p,x)$
approaches the limit $g^B_{\infty}(p,x) \equiv [1-2p(1-p)x -
\sqrt{1-4p(1-p)x}~]/2p^2x$, which can be expanded in powers of
$x$.  On integrating the coefficients of this expansion with
respect to
$p$ we obtain $\Pi^B(s) = {1/s(2s+1)}$ for $s\leq N$, which implies
an
asymptotic exponent $\gamma_1^B = 2$.

Next we consider Definition A, and show that $\Pi^A(s)\sim 1/s$ for
large $s$, so that $\gamma_2^A =1$.  For large $N$ and large $s$,
it
follows from general theorems\cite{Harris} that for $p>{1/over 2}$
\begin{equation}
\label{e.10}
P_N^A(p,s) \sim s_0^{-1}\exp\left[-C(s/s_0)^{\gamma(p)}\right], ~~~
s_0=(2p)^N.
\end{equation}
Here $s_0$ is the average avalanche size (number of open alveoli)
in
the generation $N$, $\gamma(p)$ is a continuous function of $p$ for
$p<1$, and $C$ is a positive number with a weak dependence on
$(s/s_0)$.
For $p \leq {1\over 2}$, $s_0$ decays exponentially with $N$. Thus,
for
$p \leq {1\over 2}$ the coefficients $P_N^A(p,s)$ for large $s$
become
negligibly small and do not contribute to the integral (\ref{e.6}).
In
contrast, for $p >{1\over 2}$ the probability of a non-zero
avalanche in
definition A --- which is equal to the sum of all the $P_N^A(p,s)$
with
$s \geq 1$ --- is finite when $N \rightarrow
\infty
 $ and equal to $(2p-1)/p^2$. This quantity should be used as
normalizing constant in the equation (\ref{e.10}).  Integrating
equation
(\ref{e.10}) with respect to $p$ from ${1\over 2}$ to 1 with the
help of the
saddle point approximation, we find

\begin{equation}
\label{e.11}
\Pi^A(s) \sim {1\over sN}(1-s^{-1/N}) ~~~ ~~~ [ N\ll s\ll 2^N],
\end{equation}
so $\gamma_2^A =1$.  If we expand Eq.~(\ref{e.11}) for small $s$,
we
find $\Pi^A(s) \sim \ln s/(sN^2)$; hence we expect to find an
effective exponent that is smaller than the asymptotic value
$\gamma_2^A
=1$, and indeed our simulations give $\gamma_2^A =0.9$ (Fig.2). 
For
very large $s$, comparable with $2^N$, the saddle point
approximation is
no longer valid and we observe (Fig.2) the ``kink'' near the end of
the
distribution\cite{text2}; Eq.~(\ref{e.11})  also holds for
definition
A, so $\gamma_2^B=1$.  Note that Eq.~(\ref{e.11}) is
valid for trees with any coordination number.

Having derived the above exponents analytically, we next examine
their
``universality'' by discussing how deviations from the assumptions
made
in the model may affect the scaling exponents. (i) The first
assumption
(which matters only for Definition B) is that we neglect the fact
that
in the lung the length $\ell$ and the radius $r$ and hence the
volume of
the airways depend on the generation number $i$\cite{Horsf}.
Previously, we modeled this generational dependence such that
$\ell_{i+1}=\ell_i/0.9$, and $r_{i+1}=r_i/0.86$\cite{Suki94}, where
the
scaling factors (0.9 and 0.86) arise from morphological
data\cite{Horsf}.   This exponential dependence should not affect
the
scaling behavior, an expectation we verified by simulations.  (ii)
The
second assumption, that the distribution of $P_{ij}$ is uniform,
matters
for both definitions.  Unfortunately, direct experimental data on
the
distribution of $P_{ij}$ in the lung are not available.  However,
even
if the distribution is not uniform, but normal or exponential, the
scaling exponents will not be influenced as long as the values of
$P_{ij}$ are not correlated.  Correlations among $P_{ij}$ have not
been
reported.  (iii)  While the assumption of a uniform distribution is
physiologically reasonable, it is also possible that there is a
weak
generational dependence of $P_{ij}$\cite{distr} which can reduce
the
scaling region and/or change the value of the exponents. A stronger
generational dependence of $P_{ij}$ in which the mean of $P_{ij}$
as a
function of $i$ increases from the root to the bottom by at least
a
factor of 10 will, however, break down the scaling
behavior\cite{distr2}. As a consequence, the very existence of
scaling
exponents found in experimental data\cite{Suki94} provides indirect
evidence that the distribution of $P_{ij}$ does not have any
generational dependence.

In summary, we have studied a statistical mechanical model of the
distribution of the {\it first\/} avalanches during lung inflation. 
Our
main result is an analytically-soluble model which, compared to the
more
realistic model of Ref.\cite{Suki94}, permits exact calculation of
the
scaling exponents with the avalanche size defined either as the
number
of alveoli (Definition A).  or number of opened airways (Definition
B).
We have found that the tree-like structure of the airways with the
simplest
assumptions concerning opening theshhold pressures is sufficient to
explain the existence of power-law distributions observed
experimentally\cite{Suki94}.  Finally, the fact that the size
distribution of the
first avalanches follows a power law suggests that high pressures
for at
least short periods of inspiration might be necessary to open up
larger
alveolar volumes. Thus, our results may also find important
applications
in the design of appropriate waveforms for artificial ventilation
of
patients who suffer from substantial airway closure and alveolar
collapse.


We thank P.~Ch.~Ivanov,
R. Sadr, A.~Shehter, K. Sneppen, and especially M. Wortis for very
helpful
comments, and NSF grant BES-9503008 and OTKA 2675 for financial
support.

\begin{figure}\narrowtext
\caption{Schematic diagram of the airways represented by a
branching tree.  The airways are labeled by a generation number
$(i=0,...,N)$ and a column number $ (j=0,...,2^i-1)$. A threshold
pressure $0<P_{ij}<1$ chosen from a uniform distribution is also
assigned to each airway.  }
\label{fig1}
\end{figure}

\begin{figure}\narrowtext
\caption {
Double logarithmic plot of the avalanche size distributions
$\Pi(s)$,
obtained by computer simulation on a Cayley tree of 12 generations.
Shown are data obtained for $10^8$ realizations for each of the two
avalanche size definitions discussed in the text, Definition A
(closed
circles) and Definition B (open circles).  Also shown, for
comparison,
are the exact results obtained using the generating function
approach
described in the text.  Again, both definitions are shown:
Definition A
(solid line) and Definition B (dotted line).} \label{fig2}
\end{figure}
\end{multicols}


\begin{references}
\bibitem{note2} The time required to open an airway is well under
0.05~s for the smaller airways
\protect\cite{Gav90}.  Since the inflation time in the
experiments was 80~s \protect\cite{Suki94}, the process of opening
an
airway can be considered to be instantaneous.
\bibitem{text2}
For very small $s<<N$ approximation (\ref{e.10}) is also incorrect. 
For
example, one can show that $\Pi^A(1) \sim 1/N^3$.
\bibitem{distr2} If $r_{i+1} = r_i/0.86$ as in Ref. \cite{Suki94},
then in a 14-generation tree $P_{ij}$ will increase by no more than
a
factor of 10. Since it was suggested that $P_{ij} \sim
1/r_i$\cite{Gav90}, we modeled the generational dependence of
$P_{ij}$
so that the mean of $P_{ij}$ depended on $i$ according to
$P_{i+1,j} =
P_{ij}/0.86$.  Simulations in a 12-generation tree showed that
using
Definition A, the scaling completely breaks down, and with
Definition B,
the scaling region is significantly reduced to about one decade.
\end{references}
 \end{document}